\documentclass[
10pt,aps,prl,twocolumn,showpacs,superscriptaddress,nofootinbib,nobibnotes,longbibliography,floatfix]{revtex4-2}

\usepackage{amsmath}
\usepackage{graphicx} %
\usepackage[utf8]{inputenc}
\usepackage[T1]{fontenc}
\usepackage{enumitem}
\usepackage{bm}
\usepackage{mathtools,amsmath,amssymb,amsfonts,graphicx,tensor,csquotes,accents, commath,chngcntr,siunitx}
\usepackage[dvipsnames]{xcolor}
\usepackage{hyperref}
\hypersetup{colorlinks=true, citecolor=MidnightBlue,
            linkcolor=MidnightBlue, urlcolor=MidnightBlue, linktocpage=true}
\usepackage[normalem]{ulem}
\usepackage{tensor}
\usepackage{leftindex}
\usepackage{caption}
\usepackage{subcaption}
\usepackage{amssymb} %
\usepackage{orcidlink}
\usepackage{pifont}
\usepackage{lmodern}
\usepackage{cancel}

\usepackage{verbatim}

\usepackage{float}

\usepackage{array}

\usepackage{xcolor}
\usepackage{multirow}
\usepackage{tabularray}
\UseTblrLibrary{booktabs}

\newcommand{\sissa}{SISSA, Via Bonomea 265, 34136 Trieste, Italy \&  INFN Sezione di Trieste}

\begin{document}

\title{When vacuum breaks: a self-consistency test for astrophysical environments in extreme mass ratio inspirals}

\author{Lorenzo Copparoni}
\email{lcopparo@sissa.it}
\affiliation{\sissa}
\affiliation{IFPU - Institute for Fundamental Physics of the Universe, Via Beirut 2, 34014 Trieste, Italy}

\author{Rohit S. Chandramouli}
\email{rchandra@sissa.it}
\affiliation{\sissa}
\affiliation{IFPU - Institute for Fundamental Physics of the Universe, Via Beirut 2, 34014 Trieste, Italy}

\author{Enrico Barausse}
\email{barausse@sissa.it}
\affiliation{\sissa}
\affiliation{IFPU - Institute for Fundamental Physics of the Universe, Via Beirut 2, 34014 Trieste, Italy}

\date{\today}
 
\begin{abstract} 
Gravitational-wave signals are typically interpreted under the vacuum hypothesis, i.e. assuming negligible influence from the astrophysical environment. This assumption is expected to break down for low-frequency sources such as extreme mass ratio inspirals (EMRIs), which are prime targets for the Laser Interferometer Space Antenna (LISA) and are expected to form, at least in part, in 
dense environments such as Active Galactic Nuclei or dark-matter spikes and cores. Modeling environmental effects parametrically is challenging due to the large uncertainties in their underlying physics. We propose a non-parametric test for environmental effects in EMRIs, based on assessing the self-consistency of vacuum parameter posteriors inferred from different portions of the signal.
Our results demonstrate that this test can reveal statistically significant inconsistencies from vacuum signals --
arising from e.g. incomplete modeling, environmental effects or deviations from General Relativity --
without introducing additional parameters or assumptions about the underlying physics.
\end{abstract}

\maketitle

\emph{Introduction.}---
With the latest release of gravitational-wave (GW) data by the LIGO-Virgo-Kagra (LVK) collaboration, more than 200 GW events have been identified, with a few black hole (BH) binaries exceeding $\sim 100 M_{\odot}$~\cite{LIGOScientific:2025slb,LIGOScientific:2025rsn}.
The next decade of GW astronomy will potentially see more massive sources through space-based detectors such as the Laser Interferometer Space Antenna (LISA)~\cite{Amaro-Seoane:2017,LISA:2024hlh}.
One of the most interesting  sources that LISA is expected to detect is extreme mass ratio inspirals (EMRIs), e.g. binaries of a stellar mass BH orbiting  a supermassive one~\cite{Barack:2003fp,drasco_gravitational_2006,Babak:2017tow}.
Because of their extreme mass ratio, these sources perform  up to $\sim 10^5$ orbital cycles in the LISA band, thus requiring highly accurate waveforms to extract astrophysical parameters~\cite{LISAConsortiumWaveformWorkingGroup:2023arg,Khalvati:2024tzz}.
At least a fraction of EMRIs are expected to form in gas-rich astrophysical environments (e.g. Active Galactic Nuclei -- AGNs)~\cite{derdzinski_-situ_2023}. EMRIs may also form in dense dark-matter environments (e.g. spikes in cold particle  dark matter scenarios~\cite{Bertone:2004pz,Barausse:2014tra,Coogan:2021uqv,Mitra:2025tag}, or bosonic cores for ultralight dark matter~\cite{Baumann:2018vus,Berti:2019wnn}). These dense environments  
can  leave a detectable imprint on their  GW signals through physical effects such as migration torques from accretion disks, gas or dark-matter accretion and dynamical friction, resonances, or even direct gravitational pulls from neighboring matter/objects~\cite{Barausse:2007dy,Yunes:2010sm,Gair:2010iv,Kocsis:2011dr,Barausse:2014tra,Baumann:2018vus,Berti:2019wnn,derdzinski_probing_2019,Deme:2020ewx,Caputo:2020irr,Coogan:2021uqv,Chandramouli:2021kts,Speri:2022upm,Duque:2024mfw,Copparoni:2025jhq}. 
Using vacuum EMRI waveform templates can thus potentially lead to  incorrect inference of the source parameters~\cite{Speri:2022upm}.

Already in vacuum, EMRI waveforms have rich signal morphology and complexity across parameter space~\cite{Barack:2003fp,drasco_gravitational_2006,hughes_adiabatic_2021,Drummond:2023wqc,chapman-bird_fast_2025,speri_fast_2024}.
 Conversely, the  literature studying signatures of environmental effects in EMRIs typically makes simplifying assumptions about the astrophysical medium. For instance, migration torques, which are among the dominant effects, 
are  typically studied in stationary, thin and radiatively efficient accretion disks within Newtonian gravity~\cite{levin_starbursts_2007,Kocsis:2011dr,Barausse:2014tra,derdzinski_probing_2019,Duque:2024mfw}.
With these  approximations,  disk-driven migration  can be modeled as a power law correction to the vacuum GW flux~\cite{Kocsis:2011dr,Barausse:2014tra, Speri:2022upm}.
Later studies confirmed that this power-law model remains robust even under the stochastic migration torques seen in simulations~\cite{Copparoni:2025jhq}.
However, the physics underpinning accretion disks in AGNs is still largely uncertain when it comes e.g. to disk geometry, density and angular momentum profile, viscosity, the role of turbulence and magnetic fields, etc.~\cite{Kocsis:2011dr,Barausse:2014tra,Speri:2022upm,Nouri:2023}. Even larger uncertainties characterize the interaction of EMRIs with dark matter, whose density profile and very nature are unknown, or the interaction with third bodies (due to their transient nature).
Relativistic effects in the interaction between the source and the environment can also become important for EMRIs, and may
lead to significant phase differences in the gravitational signals~\cite{Barausse:2007ph,Vicente:2025gsg,HegadeKR:2025dur,HegadeKR:2025rpr,Duque:2025yfm}. Finally, 
multiple environmental effects (e.g. migration and accretion in the case of disks)
and even possible violations of General Relativity (GR) may be simultaneously present, thus complicating the signal analysis and leading to degeneracies~\cite{Speri:2022upm,kejriwal_impact_2024}. 
Due to these technical difficulties and physical uncertainties, a natural question is whether EMRI GW data can robustly reveal environmental effects, if the latter are (at least partially) mismodeled.

In this \emph{Letter}, we show that one can indeed test for the presence of 
astrophysical environments, or even deviations from General Relativity (GR), using {\it only} vacuum EMRI waveform templates, i.e. assuming no specific model for effects 
that go beyond vacuum GR waveforms. 
Environmental effects are expected to be generally stronger during the early inspiral (i.e., at lower  frequencies)~\cite{Barausse:2014tra} and are  suppressed, in relative terms, closer to the plunge/merger.
If that is the case, the last portion of the signal will be well described by a vacuum EMRI waveform, while inconsistencies/biases  arising from neglecting environmental effects will build up from the early inspiral.
By constructing data segments of increasing duration, we develop a self-consistency test that compares the vacuum EMRI parameter posteriors obtained from each  segment.
When the self-consistency test fails, our method indicates the presence of missing, unmodeled effects in the vacuum EMRI template. 
We apply this test to a fiducial EMRI signal containing an environmental effect (disk-driven migration), and quantify the significance of the inconsistency in the posteriors from different observation durations. 
We do so by computing the mismatch and relative systematic error between the estimated maximum likelihood points. 
We also estimate the probability that the inconsistency between the posteriors from different signal portions is due to the  noise realization.
In our fiducial system, we show that for observations of more than two years, the presence of an environmental effect  can be robustly identified, irrespective of the noise realization.
Throughout the paper, we use $M$ for primary mass, $\mu$ for secondary mass, and geometric units with $G=1=c$.

\emph{Waveforms and data analysis methods.}---
We compute EMRI waveforms using the time-domain multivoice method~\cite{drasco_gravitational_2006,hughes_adiabatic_2021} implemented in the Fast EMRI Waveform (\texttt{few}) package~\cite{chapman-bird_fast_2025, speri_fast_2024, katz_fastemriwaveforms_2021}.
Specifically, these waveforms capture fully relativistic equatorial Kerr orbits evolving adiabatically within the self-force formalism~\cite{Hinderer:2008dm,hughes_adiabatic_2021}.
To generate EMRI waveforms containing an environmental effect, we add suitable contributions to the adiabatic fluxes, which in turn modify the orbital evolution and thus the waveform.
For simplicity, we implement  contributions from disk-driven migration torques as computed for thin disks  within Newtonian theory~\cite{Kocsis:2011dr,Barausse:2014tra,Speri:2022upm}. For circular prograde orbits, the angular momentum flux then takes the form 
$\dot{L} = \dot{L}_{\rm GW} A (r/r^*)^n$, 
where $\dot{L}_{\rm GW}$ is the Newtonian point-particle GW flux~\cite{Peters}, $r^* = 10 M$ is a characteristic separation, with $A$ and $n$ mapping to different accretion disk models~\cite{Kocsis:2011dr,Barausse:2014tra,Speri:2022upm}.
As proof-of-principle of our self-consistency test,
we  focus here on $n=8$, corresponding to  Type-I migration~\cite{Yunes:2011ws,tanaka_okada} in a Shakura-Sunayev $\alpha$-disk~\cite{shakura_sunyaev}.
We choose $A=1.92\cdot10^{-5}$, as in the analysis
by~\cite{Speri:2022upm}, which is consistent, but not limited to,  with an accretion Eddington ratio $f\textsubscript{Edd} = 0.0005$ and a viscosity parameter $\,\alpha = 0.03$.
In the Supplemental Material, we also extend this analysis to other environmental effects (by varying $n$) and different source parameters, to further demonstrate the wider applicability of our self-consistency test.

We inject such an EMRI signal into the time delay interferometry (TDI)~\cite{Tinto:2003vj} channels of LISA using \texttt{fastlisaresponse}~\cite{Katz:2022yqe}. 
Using vacuum EMRI  templates, we  perform Bayesian parameter estimation of $\boldsymbol{\theta} = \{ \log_{10} M/M_\odot$, $\log_{10} \mu/M_\odot$, $a$, $r_0/M$, $D_L/\mbox{Gpc}$, $\cos \theta_S$, $\phi_S$, $\cos \theta_K$, $\phi_K$, $\Phi_{\phi_0}\}$. 
Note that $a$ is the dimensionless spin of the primary, $r_0/M$ and $\Phi_{\phi_0}$ are the initial separation and orbital phase, $D_L/\mbox{Gpc}$ is the luminosity distance, $\theta_S$ and  $\phi_S$ are the azimuthal and longitudinal sky position in the solar system barycenter frame, and $\cos \theta_K$, $\phi_K$ are the primary spin orientation relative to the line of sight.

For the injection, we use  vacuum EMRI parameters $\boldsymbol{\theta}_{\rm inj} = \{6.0$, $1.69897,$ $0.9,$ $14.382,$ $3.16,$ $0.751,$ $0.236,$ $0.555,$ $0.628,$ $0.720\}$, which for a $3~\mbox{yr}$ observation \footnote{The injection value for $r_0/M,\,\Phi_{\phi_0}$ are the initial separation and phase corresponding to the $3~\mathrm{yr}$ observation. These injected values change with the observation window.} yields a signal-to-noise ratio (SNR) of $100$ and plunges about $12600~\mathrm{s}$ before the end of the observation window.
We note that this
is  reasonable SNR based
on astrophysical population models~\cite{Babak:2017tow,Pan:2021ksp,Pan:2021oob,Kejriwal:2025jao}. In the Supplemental Material, we also consider a different EMRI system with a lower SNR, to demonstrate how our test performs for weaker sources.
We use uniform priors on all parameters except the luminosity distance, for which we use $p(D_L)\propto D_L ^2$.
Assuming stationary instrumental Gaussian noise, we write the joint log-likelihood of the TDI channels  (up to an additive constant) as~\cite{Marsat:2020rtl}:
\begin{align}
 \log \mathcal{L}(d | \boldsymbol{\theta}) = -\dfrac{1}{2}\sum_{i} \left ( d_i - h_i(\boldsymbol{\theta}) \left \lvert \right. d_i - h_i(\boldsymbol{\theta}) \right)_i, \label{eqn:logL}
\end{align}
where $d_i$ and $h_i(\boldsymbol{\theta})$ are the data 
and the waveform model, respectively, in TDI channel $i \in \{A, E\}$.
The T channel is not included in our analysis, as it is an antisymmetric TDI combination and is thus mostly dominated by noise~\cite{Marsat:2020rtl}. 
The matched filter inner product in channel $i$ between time-series $u_i(t)$ and $v_i(t)$ is defined as
\begin{align}
    \left ( u_i \left \lvert \right. v_i \right)_i = 4 \mathrm{Re} \int_0^{\infty} \dfrac{\mathrm{d}f}{S_n^i(f)} \tilde{u}_i(f)^* \tilde{v}_i(f), \label{eqn:inner-product}
\end{align}
where $S_n^i(f)$ is the  one-sided noise power spectral density (PSD) of TDI channel $i$, with $\tilde{u}_i(f)$ and $\tilde{v}_i (f)$ the Fourier transforms of the respective time series.
We obtain corresponding posterior samples using the parallel tempered Markov Chain Monte Carlo sampler implemented in \texttt{eryn}~\cite{karnesis:2023ras} (see Supplemental Material for more details). 
We use the \texttt{scirdv1} version of the LISA PSD with second generation TDI~\cite{Tinto:2022zmf}, to which we add the confusion noise from unresolved galactic binaries~\cite{cornish_galactic_2017}. 
We also consider specific Gaussian noise realizations in the injected data, to assess their impact on the parameter estimation results.

\begin{figure}[t]
    \centering
    \includegraphics[width=0.45\textwidth]{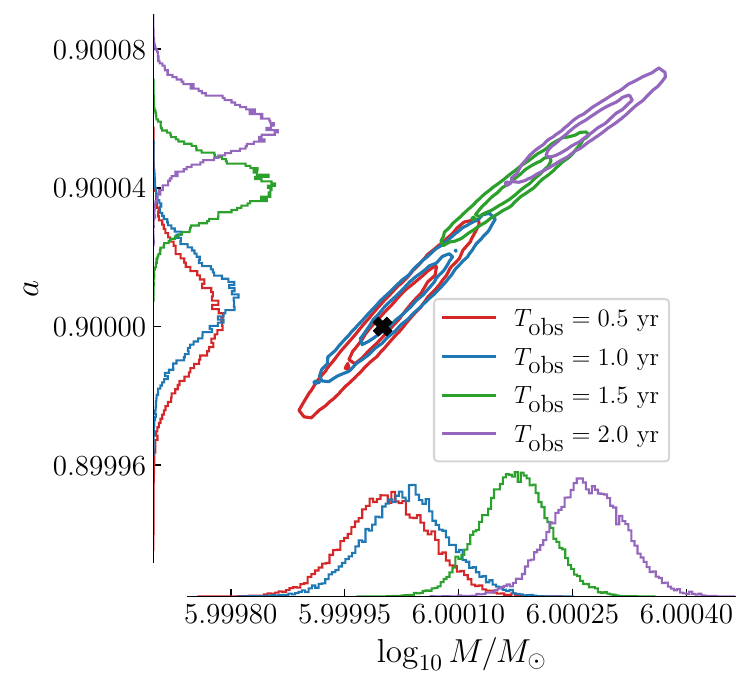}
    \caption{Two dimensional marginalized posteriors (with contours at $50\%$ and $90\%$ confidence level) for the primary mass $M$ and spin $a$, for an injection with disk migration torques but with   vacuum EMRI templates, for
    different observation durations.
    Significant inconsistencies in the posteriors reveal the neglected environmental effect.
    For clearness sake we indicate the true value of the parameters with a black cross.
    }
    \label{fig:posterior_consistency}
\end{figure}

\begin{figure*}[t]
    \centering
    \includegraphics{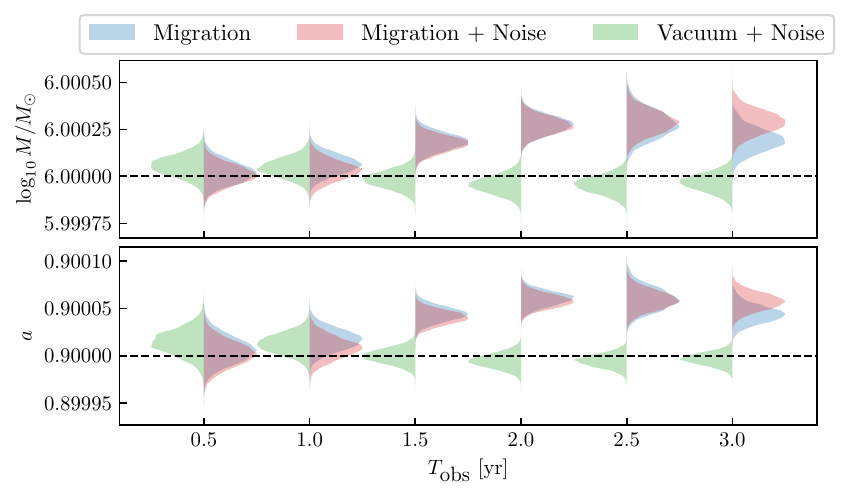}
    \caption{One dimensional marginalized posteriors of $\log_{10} M/M_\odot,\, a$ for different injections, namely: a vacuum EMRI with noise (green); a migration-affected EMRI with (red) and without (blue) noise. The inconsistency in the posteriors due to the environment can be distinguished from the effect of the noise
realization.
    }
    \label{fig:D1_violinplot}
\end{figure*}

\emph{Inconsistent inference across observation duration.}---
To perform a self-consistency check of the vacuum EMRI templates, we compare the parameter posteriors resulting from Bayesian inference, with different signal durations.
Specifically, we consider  signal segments  with $T_{\rm obs} \in \{0.5,1,1.5,2,2.5,3\}~\mathrm{yr}$, all ending at the same time (right after the plunge).
Since most of the SNR is in the last part of the waveform, this ensures that we  obtain informative posteriors for the vacuum EMRI parameters for all observation times.
Given the injected EMRI signal with Type-I $\alpha$-disk  migration torques, we show  in Fig.~\ref{fig:posterior_consistency} the inferred one-dimensional and two-dimensional marginalized posteriors of the primary mass $M$ and dimensionless spin $a$, for different $T_{\rm obs}$.
With increasing $T_{\rm obs}$,  the inferred posteriors become more and more inconsistent.
In more detail, the posteriors obtained with the vacuum EMRI model for \( T_{\rm obs} = 2\,\mathrm{yr} \) show minimal overlap with those for \( T_{\rm obs} = 0.5\,\mathrm{yr} \).
This reveals the presence of a missing effect in the model---disk-driven migration in this example---that has accumulated significantly between \( T_{\rm obs} = 0.5 \) and \( 2\,\mathrm{yr} \) of observation.

To contrast the role of instrumental noise with that of the environmental effect, we consider a vacuum EMRI injection with a Gaussian noise realization, and compare it with  injections featuring disk-driven migration with and without such a noise realization.
In Fig.~\ref{fig:D1_violinplot}, we compare the resulting one-dimensional marginalized posteriors for $\log_{10} M/M_\odot,\,a$ from these different injections (see Supplemental Material for a full corner plot).
When injecting a vacuum EMRI signal, we observe that the (green) posteriors are statistically self-consistent, with shifts induced by the noise realization that have no significant trend with observation duration.
However, the (blue) posteriors resulting from the environment-affected EMRI injection without noise clearly show significant inconsistencies growing with the observation duration.
Note that given the high SNR,
the posteriors obtained from this noiseless injection are a good approximation for the ensemble average (over noise realizations) of the posteriors obtained with noisy injections~\cite{Nissanke:2009kt,Vallisneri:2011ts}.
Importantly, we also observe good agreement between these `average'
posteriors and those obtained for en environment-affected EMRI with a specific noise realization (red).
We find that for short observation durations $T_{\rm obs} \leq 1~\mbox{yr}$, the inconsistency due to the missing environmental effect is comparable with the effect of the instrumental noise realization.
For longer observation durations, the inconsistency is apparent and  becomes distinguishable from the effects of the noise realization.

We will now quantify the statistical significance of the inference inconsistency, shown in Fig.~\ref{fig:posterior_consistency} and Fig.~\ref{fig:D1_violinplot}, using both a \emph{mismatch} criterion and a \emph{relative systematic bias}, 
which we describe below.
For a given waveform model (including the LISA response) $h (\boldsymbol{\theta})$, the match $ \mathcal{M}(\boldsymbol{\theta}_1,\boldsymbol{\theta}_2)$ quantifies the distance between two points $\boldsymbol{\theta}_1$ and $\boldsymbol{\theta}_2$ in parameter space, with the mismatch given by $1-\mathcal{M}(\boldsymbol{\theta}_1,\boldsymbol{\theta}_2)$. 
In a given channel $i$, the match is expressed as
\begin{align}
    \mathcal{M} (\boldsymbol{\theta}_1,\boldsymbol{\theta}_2) = \max_{\delta t, \delta \phi }\dfrac{\left( h_i (\boldsymbol{\theta}_1) \left \lvert \right. h_i (\boldsymbol{\theta}_2) \right)_i}{ \| h_i (\boldsymbol{\theta}_1)\|_i \| h_i (\boldsymbol{\theta}_2)\|_i},\label{eqn:match}
\end{align}
where $\| u_i \|^2_i = (u_i | u_i)_i$ defines the norm squared with the inner products defined in Eq.~\eqref{eqn:inner-product}.
Following~\cite{Moore:2018kvz}, the match is maximized over  overall phase and time shifts, which are needed to align the waveforms.
When computing the match across all channels, denoted as $\mathcal{M}_{\rm net}$, we sum over each inner product as discussed in~\cite{Cutler:1994ys}. 

Using the maximum likelihood estimate (MLE) $\boldsymbol{\theta}_{\rm MLE,ref}$ obtained with $T_{\rm obs} = 0.5\,\mathrm{yr}$ as a reference, we compute the matches (using the vacuum EMRI waveform templates) to $\boldsymbol{\theta}_{\rm MLE}$ as obtained from other observation durations.
We denote these matches by $\mathcal{M}_{\rm net, T_{obs}} \equiv \mathcal{M}_{\rm net}(\boldsymbol{\theta}_{\rm MLE,ref},\boldsymbol{\theta}_{\rm MLE})$ for convenience.
In each case, to compute the match, we generate the  waveforms using an observation duration of $T_{\rm obs} = 0.5,\mathrm{yr}$, thus ensuring that their duration matches that of the reference. 
Based on~\cite{Baird:2012cu,Toubiana:2024car,Flanagan:1997kp,Lindblom:2008cm,Chatziioannou:2017tdw,Chandramouli:2024vhw}, we consider the Bayesian inference to be significantly inconsistent when the mismatch satisfies 
\begin{align}
1-\mathcal{M}_{\rm net, T_{obs}} > \dfrac{\chi^2_{D,90\%}}{2\rho_{\rm ref}^2}, \label{eqn:mismatch_criterion}
\end{align}
with $\rho_{\rm ref} = (\sum_i \| h_i (\boldsymbol{\theta}_{\rm MLE, ref}) \|_i^2)^{1/2}$ being the optimal SNR~\cite{Cutler:1994ys} of the reference waveform, and $\chi^2_{D,90\%}$ being the $90 \%$ quantile for the chi-square distribution with $D=\mathrm{dim}(\boldsymbol{\theta})$ degrees of freedom.
For $D=10$, we have $\rho_{\rm ref} = 99$ and $\chi^2_{D,90\%} = 16$, resulting in threshold mismatch of $8.3\times 10^{-4}$.
We also assess the inconsistency in the MLEs by computing the relative systematic bias as 
\begin{align}
\delta \theta^{\alpha} \equiv \dfrac{| \theta^{\alpha}_{\rm ML,ref} - \theta^{\alpha}_{\rm ML}|}{\Sigma^{\alpha}_{\rm 90\%}}, \label{eqn:relative_bias}   
\end{align}
where $\alpha$ is the parameter index, and $\Sigma^{\alpha}_{\rm 90\%}$ is the  parameter's  $90\%$ credible interval for the reference case  $T_{\rm obs} = 0.5$yr.
With a large mismatch satisfying Eq.~\eqref{eqn:mismatch_criterion}, one would typically expect significant biases in all parameters, with $\delta \theta^{\alpha}>1$.
However, $\delta \theta^{\alpha}>1$ for a particular parameter does not imply that Eq.~\eqref{eqn:mismatch_criterion} is satisfied.
For this reason, we primarily use Eq.~\eqref{eqn:mismatch_criterion} as the main quantitative metric to identify inconsistent inferences, and report the corresponding relative systematic biases for a few key intrinsic parameters such as $\log_{10} M$ and $a$.

\begin{table}[b]
\caption{\label{tab:mismatch_params}
Mismatches and systematic biases relative to the reference MLE at $T_{\rm obs} = 0.5\,\mathrm{yr}$.
The SNR for 3 yr of observation is 100.
}
\begin{ruledtabular}
\begin{tabular}{cccc}
$T_{\rm obs}$ (yr) & $1 - \mathcal{M}_{\rm net, T_{obs}}$ & $\delta \log_{10} M/M_\odot$ & $\delta a$ \\
\hline
1.0 &  $9.0\times 10^{-5}$      &   0.10    &  0.10    \\
1.5 &  $4.8\times 10^{-4}$      &   1.60    &  1.60    \\
2.0 &  $1.4\times 10^{-3}$      &   2.58    &  2.60    \\
2.5 &  $1.6\times 10^{-3}$      &   3.13    &  3.17    \\
3.0 &  $3.2\times 10^{-3}$      &   3.10    &  3.14    \\
\end{tabular}
\end{ruledtabular}
\end{table}

In Table~\ref{tab:mismatch_params}, we show the mismatches with the reference MLE computed with Eq.~\eqref{eqn:match}, along with the relative systematic biases $\delta \log_{10} M/M_\odot$ and $\delta a$ for each observation duration.
We find that for $T_{\rm obs}\leq 1.5$yr, the mismatch is below the threshold (as given in Eq.~\ref{eqn:mismatch_criterion}).
For a longer duration  $T_{\rm obs} = 2$yr, the mismatch exceeds the threshold, indicating a significant inconsistency.
For $T_{\rm obs} \geq 2.5$yr, the inconsistency is so significant that the probability that it is induced by noise is $\leq 5\times10^{-4}$, which  corresponds to $\delta \log_{10} M/M_\odot>3$ and $\delta a>3$.
Furthermore, in the Supplemental Material we  verify that our results are robust to different noise realizations.
Our quantitative analysis with Eqs.~\eqref{eqn:mismatch_criterion}--~\eqref{eqn:relative_bias} therefore confirms the  results
show in Fig.~\ref{fig:posterior_consistency}--\ref{fig:D1_violinplot}.

\emph{Discussion.}---
Detecting and discriminating environmental effects and/or deviations from GR in EMRI signals is an important challenge for LISA data analysis.
While there are efforts underway aimed at accurately modeling specific environmental effects (due e.g. to gas or dark matter), there is still a large uncertainty in  their physics. Moreover, accurate 
models for environmental effects will necessarily involve several additional parameters, which may render their implementation in the LISA global fit~\cite{Littenberg:2020bxy,Lackeos:2023eub,Littenberg:2023xpl,Deng:2025wgk,Strub:2024kbe, Katz:2024oqg} difficult in practice.
In this work, we have therefore developed a self-consistency check
solely based on vacuum GR EMRI waveforms. Our approach tests
inconsistencies/biases in the posteriors as inferred from different portions of the signal, which can reveal missing physics in the templates, due e.g. to environmental effects and/or possible beyond-GR effects.
For an injected EMRI signal affected by disk-driven migration, we showed that the parameter posteriors are significantly inconsistent, when inferred from different portions of the signal, for observations of more than two years.
 
Crucially, our model relies on the successful subtraction of louder sources in the LISA global fit. Indeed, residual power from imperfect subtraction of overlapping sources (e.g. Galactic binaries~\cite{Khukhlaev:2025xiz}) may be flagged by our test. 
In the Supplemental Material, we consider a slight mismodeling of the galactic noise, and find that it does not result in significant inconsistencies across observation times.
We will explore a more rigorous treatment of this problem in future work.
Here, we stress that our method can also be viewed as a consistency test for the ``goodness'' of the LISA global fit.

In the main text of this work, we have focused on disk-driven migration to showcase our self-consistency test.
However, in the Supplemental Material~\cite{SupMat:1} we validate our results using two additional environmental effects (entering at $-5.5$PN and $-4$PN relative to the GW flux) and for a different set of source parameters.
Additionally, we also consider an effect that contributes at $1$PN order, in contrast to the negative-PN environmental effects, to demonstrate the broad applicability of the self-consistency test.
An important data-analysis question concerns how to interpret the outcome of the self-consistency test.
If the test fails, does this necessarily imply that the neglected physical effect would become measurable once it is explicitly modeled?
Conversely, even if the test is successfully passed, could a neglected physical effect still be present but remain hidden through degeneracies with the vacuum EMRI parameters (a form of stealth bias~\cite{Cornish:2011ys,Vallisneri:2013rc})?
Answering these questions requires a more systematic investigation of the performance of our self-consistency test across a wider range of environmental effects and over a broader region of the EMRI parameter space (for example including eccentricity).
Such a study is beyond the proof-of-principle scope of this paper and is left for future work.

We also emphasize that our approach identifies failures of the vacuum model but does not, by itself, identify which effect is missing or mitigate the parameter biases that arise.
A common approach to addressing this is to introduce additional phenomenological parameters into the model, which can potentially mitigate the biases in vacuum EMRI parameters 
(for similar applications, see~\cite{Moore:2014pda,Moore:2015sza,Read:2023hkv,Owen:2023mid,Kumar:2025nwb,Mezzasoma:2025moh,Pompili:2024yec,Hoy:2024vpc}), and also help identify missing physical effects.
Finally, we note that our self-consistency test complements other methods such as the inspiral-merger-ringdown test~\cite{ghosh_testing_2016,ghosh_testing_2018} or signal reconstruction using unmodeled searches, which can also potentially be used to identify missing physical effects~\cite{Miller:2023ncs,Das:2024zib,Seo:2025dto}.

\emph{Acknowledgments.}---
We thank Laura Sberna for  helpful discussions and for reading this manuscript.
We also thank the anonymous referee for their insightful comments.
We acknowledge support from the European Union’s Horizon ERC Synergy Grant ``Making Sense of the Unexpected in the Gravitational-Wave Sky'' (Grant No. GWSky-101167314) and the PRIN 2022 grant ``GUVIRP - Gravity tests in the UltraViolet and InfraRed with Pulsar timing''.
This work has been supported by the Agenzia Spaziale Italiana (ASI), Project n. 2024-36-HH.0, ``Attività per la fase B2/C della missione LISA''. 
Computational work have been made possible through SISSA-CINECA agreements providing access to resources on LEONARDO at CINECA.MB
\bibliographystyle{apsrev4-2}
\bibliography{references}
\subsection*{Supplemental Material}
When performing Bayesian parameter estimation, we used the tempered sampler \texttt{eryn}.
For all the analyses shown in the main text, we used a mixture of `stretch moves' and `adaptive Gaussian moves', with 2 temperatures, which allowed for a thorough exploration of the parameter space.
We further checked convergence by inspecting the trace plots for each posterior chain.
For completeness, we show in Fig.~\ref{fig:full_posterior} the one dimensional and two dimensional marginalized posteriors of both the intrinsic and extrinsic vacuum EMRI parameters, for the injection with both noise and migration torques.
We do not show the initial separation $r_0/M$ and phase $\Phi_{\phi_0}$ as they are (by definition) different for each observation duration.
Each colored histogram corresponds to $T_{\rm obs} = \{0.5,1,1.5,2,2.5,3 \}$~yr, as discussed in the main text.
This supplements Fig. 1
of the main text,
where we only showed the posterior inconsistency for the primary mass and spin, for $T_{\rm obs} = \{ 0.5, 1, 1.5, 2\}$~yr.

\begin{figure}[b]
    \centering
    \includegraphics[width=0.49\textwidth]{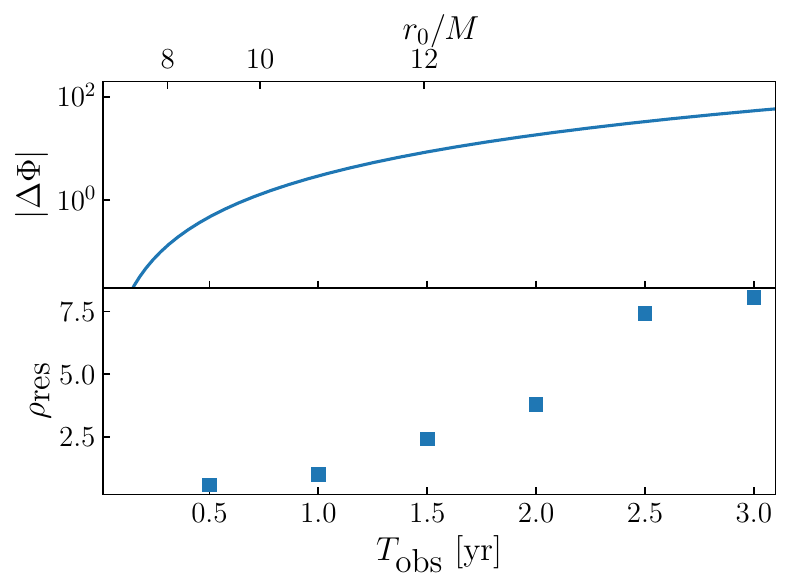}
    \caption{Top panel shows the orbital dephasing (as a function of duration and orbital separation $r_0/M$) between a noiseless  EMRI with migration torques and the corresponding vacuum case. 
    For this injection, the bottom panel shows the residual SNR obtained from the maximum likelihood vacuum EMRI waveform.
    Note that the injection  corresponds to the one in the main text.
    }
    \label{fig:res_SNR}
\end{figure}

To better understand the role of the environmental effect (disk-driven migration) on the EMRI signal, we compute the resulting orbital dephasing (relative to the vacuum case), for the same injected vacuum EMRI parameters.
In the top panel of Fig.~\ref{fig:res_SNR}, we show the  orbital dephasing as a function of both the observation duration and (equivalently) the orbital separation $r_0/M$.
Because the disk-driven migration (in a Type-I $\alpha$-disk) considered in our work is a $-8$PN effect,  Fig.~\ref{fig:res_SNR} shows a significant orbital dephasing  $\gtrsim 50$~rad for a $3$~yr observation.
Conversely, for a shorter observation of $\lesssim 1$~yr, the dephasing is $\lesssim 3
$~rad. 
As discussed in the main text, the inference inconsistencies become more significant with longer observation duration, because the environmental effects are dominant at lower frequencies.

We also report, in the bottom panel of Fig.~\ref{fig:res_SNR}, the squared residual SNR corresponding to the MLE~\cite{Lindblom:2008cm,Vallisneri:2012qq,Chandramouli:2024vhw}, given by $\rho^2 _{\rm res} \equiv - \log \mathcal{L}(d| \boldsymbol{\theta}_{\rm MLE})$, which we calculated for the  noiseless injections with migration torques.
As expected, for longer observations, an increasing amount of signal is lost in the analysis, due to the missing physics in the template.
The signal loss is never large enough to prevent detection,
but in the context of the LISA global fit,
the residual SNR from many sources may accumulate and potentially cause biases and/or missed detections~\cite{antonelli_noisy_2021}.

As a validation of our self-consistency test, we also injected a vacuum EMRI signal with noise. 
We considered the same source parameters $\boldsymbol{\theta}_{\rm inj}$  and observation durations as in the main text, and inferred posteriors on $\boldsymbol{\theta}$.
The  results are reported in Table~\ref{tab:mismatch_vac_param}.
We see that the mismatch criterion given by Eq. 4 in the main text
is always well satisfied.
Note that  noise can cause a relative systematic bias $\delta \theta^{\alpha}>1$
in the  one-dimensional marginalized posteriors, while the mismatch criterion is still satisfied.
Indeed, since the mismatch criterion takes into consideration the whole parameter space, it is a more reliable diagnostic of the presence of environmental effects than the one-dimensional posteriors.

We have checked that different noise realizations do not yield statistically different results for both the vacuum and migration injections.
In Table~\ref{tab:mismatch_noise_real}, we show the mismatches obtained with $T_{\rm obs} = 2.5~\mbox{yr}$ and $T_{\rm obs}=3.0~\mbox{yr}$ observations against the reference $T_{\rm obs} = 0.5~\mbox{yr}$ observation. We considered two different noise realizations for both the vacuum and migration injections, using the same source parameters as those described in the main text.
We find that the mismatches obtained for given $T_{\rm obs}$ are comparable
across noise realizations, and agree with Tables~\ref{tab:mismatch_vac_param} 
of supplemental
and
Table I of the main text.

Additionally, we tested whether an inconsistent inference across observations can be caused by a slight mismodeling of the noise PSD, instead of a mismodeling of a physical effect. 
We considered two vacuum injections in which the PSD used to produce the noise and the one in the likelihood calculation differ.
In the first case we assume that the acceleration component of the LISA noise curve is underestimated during the global fit.
Specifically we increase the acceleration noise component by $10\%$ in the PSD used in the noise generation compared to the one used in the data analysis pipeline.
In the second case we consider a slight mismodel in the galactic binaries foreground.
Using the empirical description for unresolved galactic binaries we simulate a situation in which the number of extracted GBs is overestimated.
Specifically we use the empirical fits introduced in~\cite{Karnesis:2021tsh} and we use an observation time which is shorter by 2 months in the PSD used for noise generation compared to the one used for parameter estimation. 
In both cases no inconsistency between the posteriors has been found and the mismatch values obtained were comparable with the ones reported in table~\ref{tab:mismatch_noise_real} for different noise realizations.
We therefore conclude that our results are robust to different noise realizations and to noise mismodeling.

 \begin{table}[t]
 \caption{\label{tab:mismatch_vac_param}
 Mismatches and systematic biases relative to the reference MLE with $T_{\rm obs} = 0.5\,\mathrm{yr}$, for a vacuum injection with noise. 
 }
 \begin{ruledtabular}
 \begin{tabular}{cccc}
 $T_{\rm obs}$ (yr) & $1 - \mathcal{M}_{\rm net, T_{obs}}$ & $\delta \log_{10} M$ & $\delta a$ \\
 \hline
 1.0 &  $5.8\times 10^{-5}$      &   0.07    &  0.04    \\
 1.5 &  $1.5\times 10^{-4}$      &   0.59    &  0.49    \\
 2.0 &  $3.2\times 10^{-4}$      &   0.92    &  0.76    \\
 2.5 &  $6.0\times 10^{-4}$      &   1.16    &  0.87    \\
 3.0 &  $4.8\times 10^{-4}$      &   0.99    &  0.74    \\
 \end{tabular}
 \end{ruledtabular}
 \end{table}

\begin{table}[t]
\caption{\label{tab:mismatch_noise_real}
Mismatch for different noise realizations, for the vacuum  and  migration injections described in the main text.
Note that the vacuum mismatches are always below the threshold discussed in the main text.
}
\begin{ruledtabular}
\begin{tabular}{c|cc|cc}
$T_{\rm obs}$ (yr) 
& \multicolumn{2}{c|}{Vacuum} 
& \multicolumn{2}{c}{Migration} \\
\cline{2-5}
& Noise 1 & Noise 2 
& Noise 1 & Noise 2 \\
\hline
2.5 & $4.8\times10^{-5}$ & $8.7\times 10^{-5}$ 
    & $1.3\times 10^{-3}$ & $1.3\times10^{-3}$ \\
3.0 & $7.0\times10^{-5}$ & $5.7\times 10^{-5}$ 
    & $1.0\times 10^{-3}$ & $8.8\times 10^{-4}$ \\
\end{tabular}
\end{ruledtabular}
\end{table}

We also explored how our self-consistency test performs for 
injections containing a different environmental effect, or corresponding to different source parameters.
For the same source parameters mentioned in the main text, which corresponds to an SNR of 100 with a $3$~yr observation, we modified the environmental effect by choosing different values of $n$ (keeping the relative torque magnitude $A$ fixed).
For event A, we set $n=4$ (that contributes at -4PN relative to the GW flux), which corresponds to environmental effects such as gas accretion and center-of-mass acceleration, or modification to gravity such as a time dependent Newton's constant~\cite{Yunes:2009bv,Yunes:2010sm,Caputo:2020irr,Toubiana:2020drf,Sberna:2022qbn}.
For events B and C, we chose $n=5.5$ (that contributes at -5.5PN relative to the GW flux), which corresponds to non-relativistic dynamical friction in a uniform collisionless medium~\cite{chandrasekhar_DF1,Ostriker:1998fa,Caputo:2020irr}.
We also chose different source parameters for event C, setting  $M =5\times 10^5~M_\odot,\, \mu = 30~M_\odot,\, a= 0.7 $, which yields an SNR of 60 for a 3yr observation.
In Tables~\ref{tab:mismatch_params_diff} and~\ref{tab:mismatch_vac_param}, we show the results of the self-consistency test for each event.

For events A and B, the posteriors become inconsistent with the reference MLE for $T_{\rm obs} \geq 2.5~\mbox{yr}$, in agreement with the accumulated dephasing relative to the vacuum template. 
Moreover, at fixed $T_{\rm obs}$, event B exhibits larger mismatches than event A.
We attribute this to the larger dephasing in event B at given $T_{\rm obs}$, since the environmental effect enters at a more negative PN order.
For event C, we find that the mismatch threshold is $2.2\times10^{-3}$, due to its lower SNR.
We observe that event C is fully incompatible with vacuum at $T_{\mathrm{obs}} = 2.0~\mbox{yr}$. 
Also note that for a longer observation time  $T_{\rm obs} = 2.5~\mbox{yr}$, the mismatch is comparable to the threshold, and then exceeds it at $T_{\rm obs} = 3.0~\mbox{yr}$.
Even though the behavior of the mismatch as a function of $T_{\rm obs}$ is not strictly monotonic for event C, unlike for events A and B, the fact that it becomes comparable to and then exceeds the threshold for $T_{\rm obs} \geq 2~\mbox{yr}$ signals an inconsistency due to using the vacuum template.

\begin{figure}[!t]
    \centering
    \includegraphics[width=0.8\linewidth]{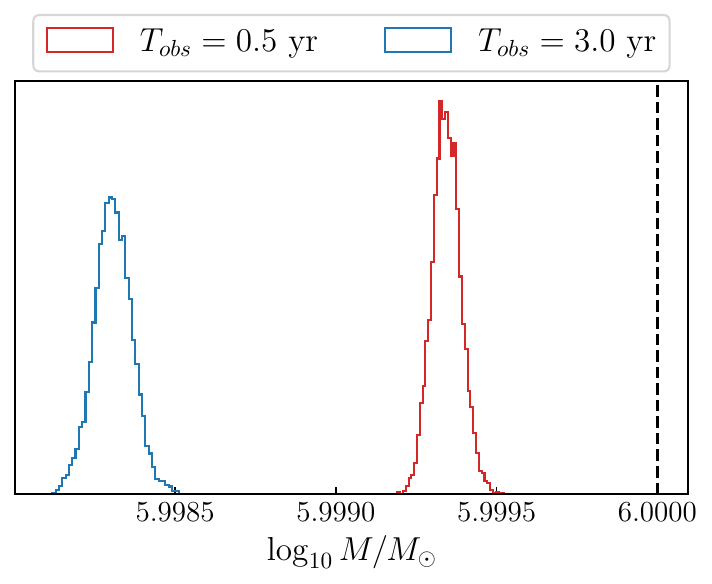}
    \caption{
    Comparison of the posterior for $\log_{10} M$ obtained from an observation time of $0.5~\mbox{yr}$ and $3.0~\mbox{yr}$.
    In black, we report the injection value.
    The injection coincides with the one in the main text for the vacuum parameters while here we have set  $A=2.0\times 10^{-4}$ and $n = -1.0$.}
    \label{fig:PN1}
\end{figure}
Since the main focus of this work is on the detection of unmodeled contribution coming from the astrophysical environment we considered negative PN modification to the gravitational fluxes.
However, positive PN modifications to the flux can instead arise due to modified gravity~\cite{Yunes_2009}.
We consider here a 1PN effect that would correspond to a propagation effect due to massive gravity~\cite{Yunes_2009}, in order to contrast against the systematics of negative PN effects.
To see the impact on the self-consistency test due to neglecting such a 1PN correction, we pick the fiducial values $A=2\times10^{-4}$ and $n=-1$ for the injection, together with the same EMRI parameters considered in the main text.
In Fig.~\ref{fig:PN1}, we show the posteriors for the primary mass obtained with a vacuum EMRI model, corresponding to a $0.5~\mbox{yr}$ (red histogram) and $3~\mbox{yr}$ (blue histogram) observation.
We see that there is significant inconsistency between the two observations, thus indicating that the self-consistency test also works for positive PN effects.
However, we also note that the inference with $0.5~\mbox{yr}$ observation is biased relative to the true injected value, unlike what we saw with negative PN order environmental effects.
This is somewhat expected as the modification becomes more significant close to merger, while in the previous cases the late inspiral modification was negligible.
The additional bias between the observation close to merger and the one with the full inspiral is due to the accumulated dephasing between
observation durations, which is why this applies to both positive and negative PN effects.

\begin{table}[b]
\caption{\label{tab:mismatch_params_diff}
Mismatches, systematic biases relative to the reference MLE at $T_{\rm obs} = 0.5\,\mathrm{yr}$ and dephasing from the vacuum trajectory, for different injections.
Events A and B have the same source parameters as the source in 
Fig. 1 in the main text,
but a different power law for the environmental torque. Event C has different source parameters ($M =5\times 10^5~M_\odot,\, \mu = 30~M_\odot,\, a= 0.7,\, \mbox{SNR} = 60.0$) and the same torque as event B.}
\begin{ruledtabular}
\begin{tabular}{cccccc}
Event & $T_{\rm obs}$ (yr) & $1 - \mathcal{M}_{\rm net, T_{\rm obs}}$ & $\delta \log_{10} M/M_\odot$ & $\delta a$ & $\Delta \Phi$\\
\hline
\multirow{3}{*}{A} & 2.0 &  $2.5\times 10^{-5}$      &   1.22   &  1.00  & 10.7 \\
& 2.5 &  $6.5\times 10^{-4}$      &   1.76    &  1.77  & 15.4 \\
& 3.0 &  $1.3\times 10^{-3}$      &   2.38    &  2.43  & 20.7 \\
\hline
\multirow{3}{*}{B} & 2.0 &  $4.6\times 10^{-4}$      &   1.55    &  1.53  & 12.8   \\
& 2.5 &  $1.2\times 10^{-3}$      &   2.69    &  2.73 & 19.9   \\
 & 3.0 &  $1.8\times 10^{-3}$      &   3.10    &  3.15 &28.7   \\   \hline          
\multirow{3}{*}{C} & 2.0 &  $3.6\times 10^{-3}$      &   1.93    &  2.00  & 34.3 \\
& 2.5 &  $2.0\times 10^{-3}$      &   1.48    &  1.51  & 53.2 \\
& 3.0 &  $2.2\times 10^{-3}$      &   1.50    &  1.54  & 76.2 \\ \end{tabular}
\end{ruledtabular}
\end{table}

\begin{figure*}[t]
    \centering
    \includegraphics[width=0.9\textwidth]{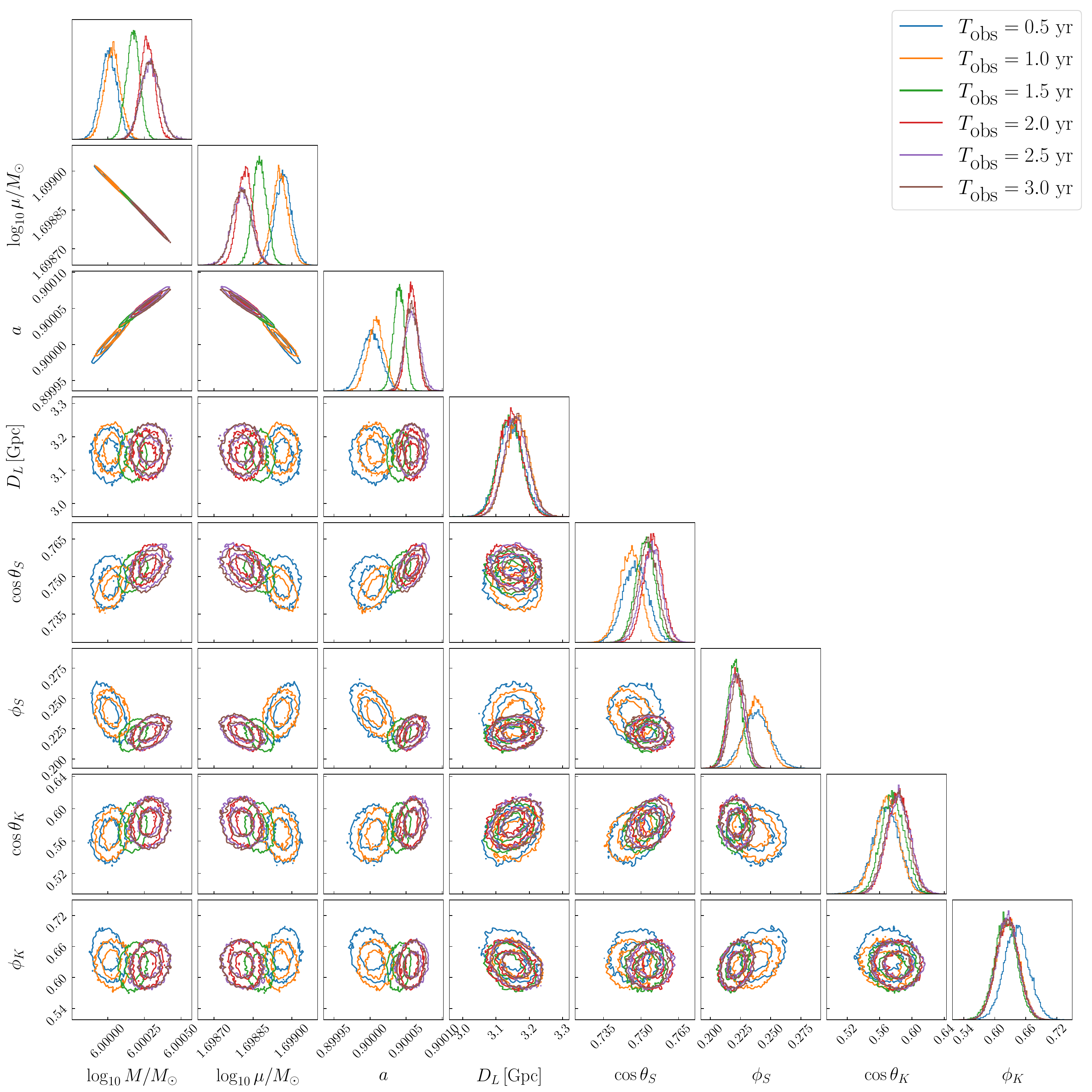}
    \caption{Posteriors for the migration + noise injection, for all the observation times considered.
    We omit the posteriors of $r_0/M$ and $\Phi_{\phi_0}$, as their true values change with  observation time.
    Note that the use of a wrong template has little effect on the extrinsic parameters posteriors, which are always consistent.
    }
    \label{fig:full_posterior}
\end{figure*}
\end{document}